\documentclass[10pt,letterpaper]{article}

\usepackage{graphicx}
\usepackage[hypertexnames=false]{hyperref}
\usepackage[dvipsnames]{xcolor}

\usepackage[numbers]{natbib}

\usepackage{bm}
\usepackage{amsfonts}
\usepackage{amsmath}
\usepackage{amssymb, amsthm, bbold, bm}
\usepackage[font={footnotesize}]{caption}
\usepackage{nameref}
\usepackage{placeins}
\usepackage{authblk}
\usepackage{graphicx}
\usepackage{subcaption} 
\usepackage{lineno}

\hypersetup{colorlinks,linkcolor={magenta},citecolor={blue},urlcolor={magenta}}

\title{Relationship between household attributes and contact patterns in urban and rural South Africa}
\author[1, 2]{Kausutua Tjikundi}
\author[3, 4]{Jackie Kleynhans}
\author[3, 4, 5]{Stefano Tempia}
\author[3, 4]{Cheryl Cohen}
\author[1]{Daniela Paolotti}
\author[1]{Ciro Cattuto}
\author[1, *]{Lorenzo Dall'Amico}
\affil[1]{ISI Foundation, 10126, Turin, Italy}
\affil[2]{University of Turin, Department of Computer Science, Turin, Italy}
\affil[3]{Centre for Respiratory Diseases and Meningitis, National Institute for Communicable Diseases of the National Health Laboratory Service, Johannesburg, South Africa}
\affil[4]{School of Public Health, Faculty of Health Sciences, University of the Witwatersrand, Johannesburg, South Africa}
\affil[5]{The Partnership for International Vaccine Initiative, Taskforce for Global Health, Atlanta, Georgia, USA}
\affil[*]{Corresponding author lorenzo.dallamico@isi.it}
\date{\today}                     
\begin{document}

\maketitle
	
\begin{abstract}
Households play a crucial role in the propagation of infectious diseases due to the frequent and prolonged interactions that typically occur between their members. Recent studies have emphasized the need to include socioeconomic variables  in epidemic models to account for the heterogeneity induced by human behavior. While sub-Saharan Africa suffers the highest burden of infectious disease diffusion, few studies have investigated the mixing patterns in the countries and their relation with social indicators. This work analyzes household contact matrices measured with wearable proximity sensors in a rural and an urban village in South Africa. Leveraging a rich data collection describing additional individual and household attributes, we investigate how the household contact matrix varies according to the household type (whether it is composed only of a familiar nucleus or by a larger group), the gender of its head (the primary decision-maker), the rural or urban context, and the season in which it was measured. We show the household type and the gender of its head induce differences in the interaction patterns between household members, particularly regarding child caregiving, suggesting they are relevant attributes to include in epidemic modeling.
\end{abstract}


\section{Introduction}

Close-range proximity interactions are the main driver of infectious disease transmission \cite{wallinga1999perspective, anderson1992infectious}. Quantifying these interactions and understanding their determinants is thus necessary to design optimal epidemic mitigation strategies. Contact matrices describe detailed data on social mixing patterns between groups of individuals and inform epidemic models by accounting for diverse social contexts and age-specific infection risks \cite{rohani2010contact}. Age is commonly adopted to define such groups \cite{edmunds1997mixes, wallinga2006using, wallinga2006using2, mossong2008social} even if recent works called for a more holistic view, integrating socio-economic variables into epidemic models \cite{gozzi2021estimating, gauvin2021socio, buckee2021thinking, tizzoni2022addressing, zelner2022there, manna2024social, manna2024generalized, manna2024importance}. Contact matrices are commonly defined within specific contexts where the interactions occurred \cite{ajelli2014role, mistry2021inferring} -- \emph{e.g.} household, school, workplace -- to account for the diverse importance each context has on the outcome of transmission. Households, in particular, play an important role in spreading the disease, as interactions within these settings are frequent and prolonged \cite{goeyvaerts2018household, longini1982estimating} and can also act as a bridge between different contexts, such as schools and workplaces. 

Although most studies and data collections focus on high-income countries \cite{hoang2019systematic}, sub-Saharan Africa experiences the highest burden of infectious disease epidemics \cite{who2024}. 
The past decade has seen an increase in interest in studying social mixing across sub-Saharan African countries to evaluate the impact of infectious disease transmission, with diverse approaches ranging from contact diaries \cite{kiti2014quantifying, melegaro2017social, le2018characteristics, de2018identifying, kleynhans2021cross} to novel technology-driven approaches based on proximity sensors \cite{kiti2016quantifying, ozella2021using, dall2022estimating, kleynhans23association}. Findings show a high degree of heterogeneity in the observed contact patterns, with few other recurring features across studies. For instance, primary school children generally display higher contact rates compared to adults \cite{le2018characteristics, melegaro2017social, kiti2014quantifying, osei2024directly}. In Kenya \cite{kiti2014quantifying, del2021individual} and Zambia \cite{dodd2016age}, rural areas record higher contact rates compared to semi-urban regions. On the opposite, in Zimbabwe \cite{melegaro2017social} peri-urban regions display higher contact rates compared to rural ones. In South Africa \cite{kleynhans2021cross} and Mozambique \cite{kiti2025social}, people living in rural areas experience almost double the contact rates as the ones in urban areas. In these settings, most interactions occur within households \cite{kiti2016quantifying, ozella2021using, van2022social}, and intergenerational contacts are driven by multigenerational households which are more common in rural areas, as reported by studies in Kenya \cite{del2021individual}, Zimbabwe \cite{melegaro2017social}, and Malawi \cite{ozella2021using}. Community-based social mixing, such as those observed in township settings in South Africa  \cite{johnstone2011social}, and Zambia \cite{dodd2016age} are associated with high rates of tuberculosis transmission in the community.
\begin{figure}[!b]
	\raggedleft
	\begin{minipage}[t]{0.43\textwidth}
		\vspace{0pt}
		\centering
		\includegraphics[width=\linewidth]{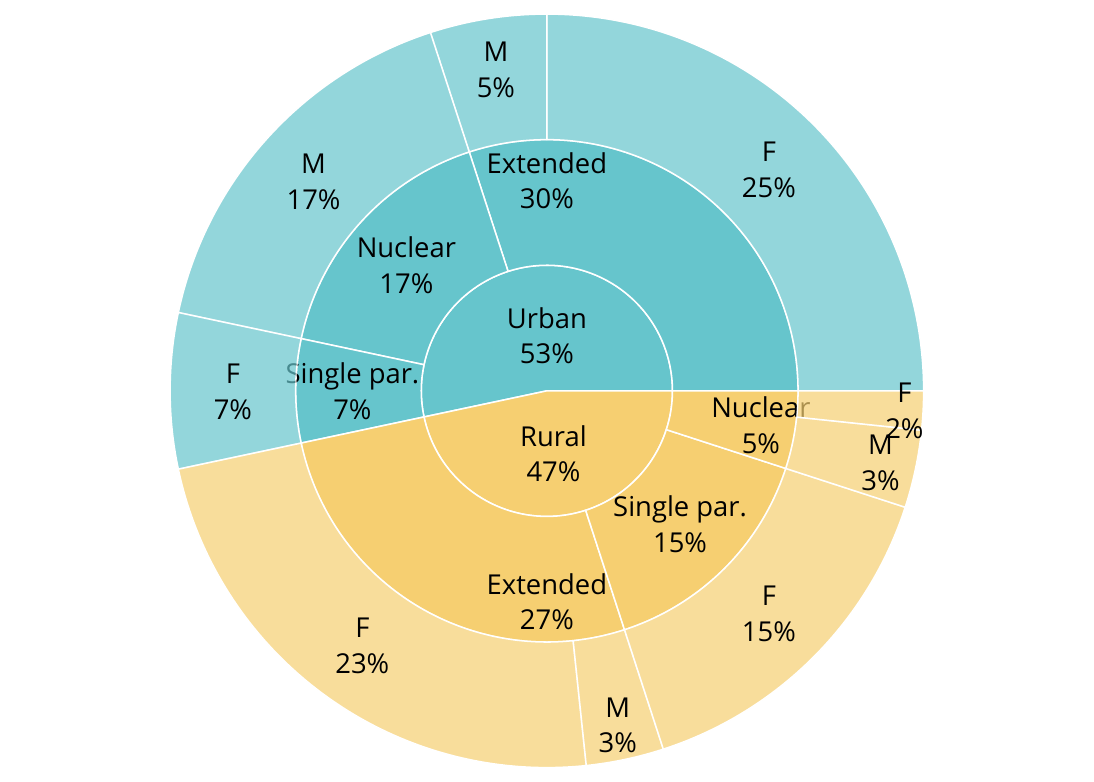}
	\end{minipage}
	\raggedright
	\begin{minipage}[t]{0.48\textwidth}
		\vspace{10pt}
		\raggedleft
		\setlength{\tabcolsep}{2pt} 
		\renewcommand{\arraystretch}{1.5} 
		\scriptsize
		\begin{tabular}{|l || c| c| c| c| c| c || c |}
				\hline
				\textbf{Classification}  & \textbf{M-Ch} & \textbf{M-Ado} & \textbf{M-Adu} & \textbf{F-Ch} & \textbf{F-Ado} & \textbf{F-Adu} & \textbf{Total}\\
				\hline
				Rural ($28$) & $32$ & $11$ & $11$ & $37$ & $16$ & $41$ & $148$\\
				Urban ($32$) & $26$ & $13$ & $24$ & $24$ & $16$ & $49$ & $152$\\
				\hline\hline
				M-headed ($17$) & $18$ & $6$ & $19$ & $12$ & $4$ & $21$ & $80$ \\
				F-headed ($43$) & $40$ & $18$ & $16$ & $49$ & $28$ & $69$ & $220$ \\
				\hline\hline
				Nuclear ($13$) & $13$ & $3$ & $13$ & $7$ & $1$ & $14$ & $51$\\
				Single-par. ($13$) & $14$ & $3$ & $0$ & $12$ & $10$ & $15$ & $54$\\
				Extended ($34$) & $31$ & $18$ & $22$ & $42$ & $21$ & $61$ & $195$ \\
				\hline\hline
				Total ($60$) & $58$ & $24$ & $35$ & $61$ & $32$ & $90$ & $300$ \\
				\hline
		\end{tabular}
	\end{minipage}
	\caption{\textbf{Summary of household and population compositions disaggregated by age and gender and according to the household classification.} \textit{Left: fraction of households disaggregated by site, household type, and gender of the household head.} The inner level indicates the fraction of households in the rural and the urban sites. The second layer categorizes households by type, while the third layer categorizes households by the gender of the household head. \textit{Right: number of individuals per age-gender for varying household grouping strategies.} The column ``Classification'' indicates the different ways of characterizing households considered in the main text. The rows that are not separated by a horizontal line indicate non-overlapping groups of households (\emph{e.g.}, Rural and Urban), while overlaps exist across lines separated by horizontal lines (\emph{e.g.}, Rural and Nuclear). We report, in parentheses, the number of households in each group. The subsequent columns indicate the number of individuals in each age group: ``M'' stands for male, ``F'' for female, ``Ch'' for children, ``Ado'' for adolescents, and ``Adu'' for adults. The column ``Total'' reports the total population in each household type (the sum over the row), while the row ``Total'' is the population per age-gender group. The rightmost bottom cell reports the total population.}
	\label{tab:hhdistr}
\end{figure}

The interaction dynamics within a household are not solely based on age, but also on the roles of its members and the overall organization of the household. In Africa, variations in this organization have been associated with poverty, urbanization, and the loosening of societal norms \cite{ibisomi201410, noumbissi20166}.
The household type (whether it is composed only of a familiar nucleus or by a larger group) and its composition (defined by age, gender, marital status, and relationships among the household members) may hence be needed to describe interactions accurately \cite{madhavan2017household, menashe2023dynamic}. Children are regarded as primary introducers of infections like influenza due to high external exposure in schools, while mothers often drive within-household transmission due to frequent caregiving interactions \cite{endo2019fine, ozella2018close}. Multigenerational households have high transmission risks for diseases like influenza and RSV, particularly affecting infants and the elderly \cite{endo2019fine, mahikul2019modeling}. In high-birth-rate settings like Kenya, larger households increase measles transmission \cite{hilton2019incorporating}, while communities composed of extended households may also be more vulnerable to larger, more severe outbreaks of infectious diseases \cite{chisholm2020model}. Similarly, in low-income settings, household composition is more intergenerational and leads to increased intra-household transmission due to more frequent interactions between older adults and younger individuals compared to high-income settings \cite{mousa2021social}. Studies confirm that household composition, rather than size alone, influences infection risk \cite{house2009household, wing2022association, mogelmose2023population}, yet disease models often assume random household mixing despite evidence of structured contact patterns \cite{goeyvaerts2018household}.  Most of these studies focused on understanding how household structure and composition influence the transmission dynamics of infectious diseases, but little effort on how structure and composition shape contact patterns relevant to the transmission of infectious diseases.

In this work, we go beyond age-only contact pattern measurements and focus on additional social attributes that may be relevant for disease spread. We consider household contact matrices measured with high resolution proximity sensors during the PHIRST study, conducted in South Africa in $2018$ \cite{cohen2021cohort, kleynhans2021cross, dall2022estimating}. These data include rich information related to the social role of individuals in the household. We investigate how household composition, headship, seasonality, and location affect the contact patterns, identifying socio-demographic attributes that may be relevant for epidemic modeling.

\section{Results}
\label{sec:res}

The data collection took place in a rural site -- Agincourt, Bushbuckridge Municipality (Mpumalanga Province) -- and an urban area -- Klerksdorp, Matlosana Municipality (North West Province) over three measurement waves with a duration between $10$ and $14$ days. The first wave was in February (South African summer), the second in April (South African autumn), and the last one in June (South African winter). In-household interactions were measured using the proximity sensors developed by the \texttt{SocioPatterns} collaboration (\href{http://www.sociopatterns.org/}{sociopatterns.org}, \cite{cattuto2010dynamics}). Section~\ref{sec:ethics} describes the ethical aspects of the study, while Section~\ref{sec:data} describes the data collection and cleaning procedures.

\medskip

We follow the study protocol that identifies households as groups of three or more people who regularly share at least two meals in the same residence at least two days per week, excluding residential institutions.
In Africa, households are diverse and shaped by various factors such as cultural diversity, economic conditions, and historical influences. South Africa is no exception, with its diverse population and the lasting influence of apartheid \cite{noumbissi20166}. To quantify the relation between the household structure and the contact patterns, we group households according to three features: i) the site, which can be urban or rural; ii) the gender of the household head, which can be male or female; iii) the household type, which can be nuclear, extended, or single-parent. The left plot of Figure~\ref{tab:hhdistr} summarizes the composition of our dataset according to the classification above. Section~\ref{sec:taxa} provides additional details on the household classification.
\begin{figure}[!b]
	\centering
	\includegraphics[width=\columnwidth]{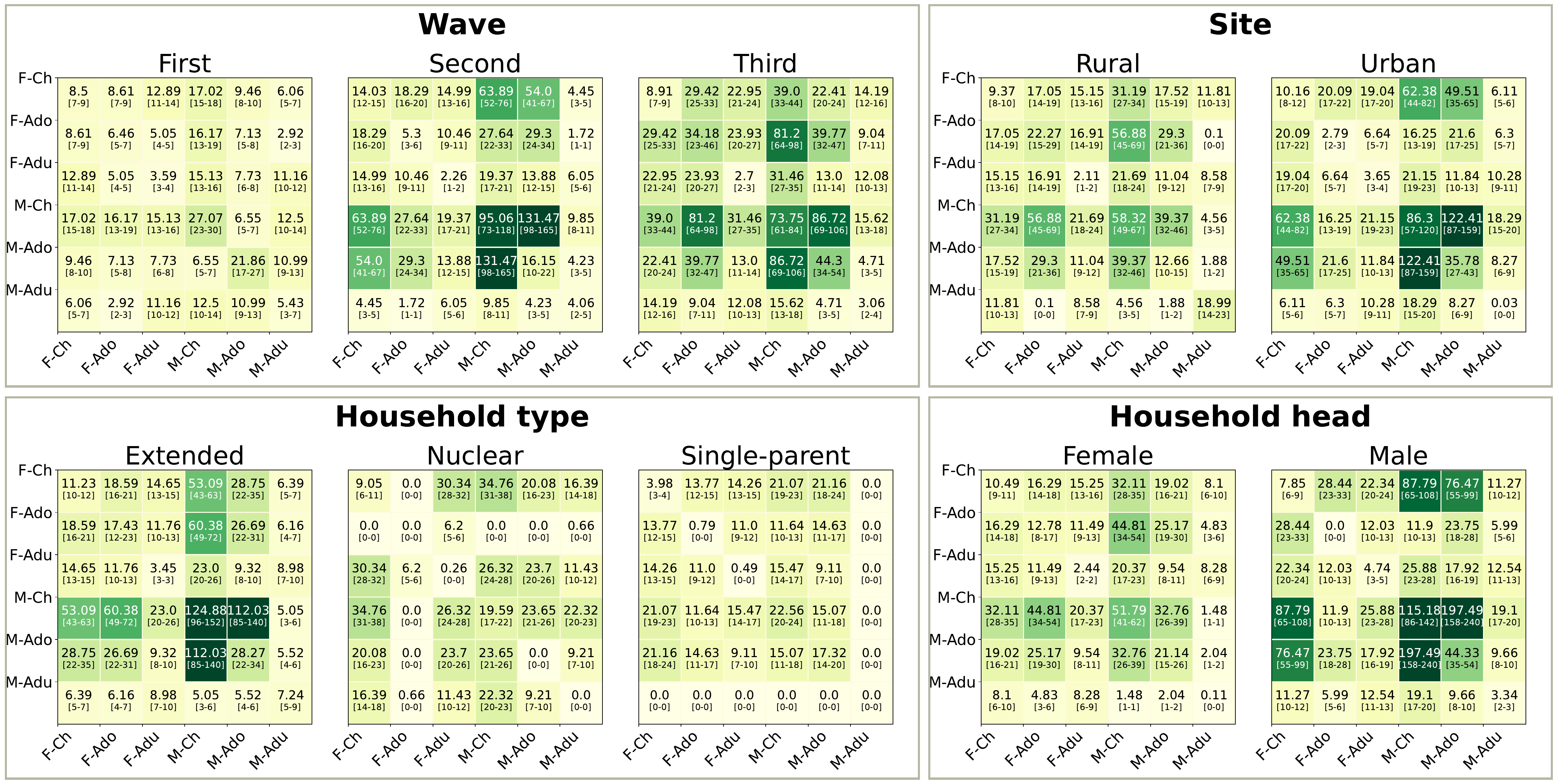}
	\caption{\textbf{Contact matrices.} Each frame represents one of the aggregations described in Section~\ref{sec:res}, plus the one obtained by grouping measurement waves. The matrices' entries denote the daily average contact duration per household by gender-age group, expressed in minutes and normalized by the population sizes, and are obtained following the procedure detailed in Section~\ref{sec:boot}. The number and color-code report the average value of the bootstrap, while the numbers in the bracket are the $10\%$ confidence interval. The population sizes for each aggregation level are reported in Table~\ref{tab:hhdistr}. The matrices are indexed by gender (``M for male, and ``F for female) and age (``Ch'' for children, ``Ado'' for adolescents, and ``Adu'' for adults).}
	\label{fig:HCM}
\end{figure}

\medskip
We analyze how household attributes impact the interaction patterns of in-household interaction rates by focusing on contact matrices describing the interaction rates between groups of individuals categorized according to their age and gender. All participants disclosed a binary gender attribute. Age categories are defined as \emph{children} for the age range $0-10$ years, \emph{adolescents} for the age range $11-18$, and \emph{adults} for individuals older than $18$ years old. We classified individuals aged 18 as adolescents rather than adults, in line with the convention that individuals typically complete their final year of schooling in the year they turn 18. The table on the right frame of Figure~\ref{tab:hhdistr} shows the population sizes disaggregated by age and gender for the different types of household classification. Figure~\ref{fig:HCM} shows the household contact matrices for each aggregation level, reporting the daily average contact duration per individual. The matrices are obtained using bootstrap sampling, and the figure reports the average value and the $10\%$ confidence interval. All matrices display a disassortative structure (assortativity index $Q$ \cite{gupta1989networks} ranging between $-0.19$ and $-0.16$), dominated by inter-generational interactions commonly observed within households \cite{mossong2008social, kiti2016quantifying, prem2017projecting}. Section \ref{sec:boot} provides further details on the definition of these contact matrices.
\begin{figure}[!t]
	\centering
	\includegraphics[width=\columnwidth]{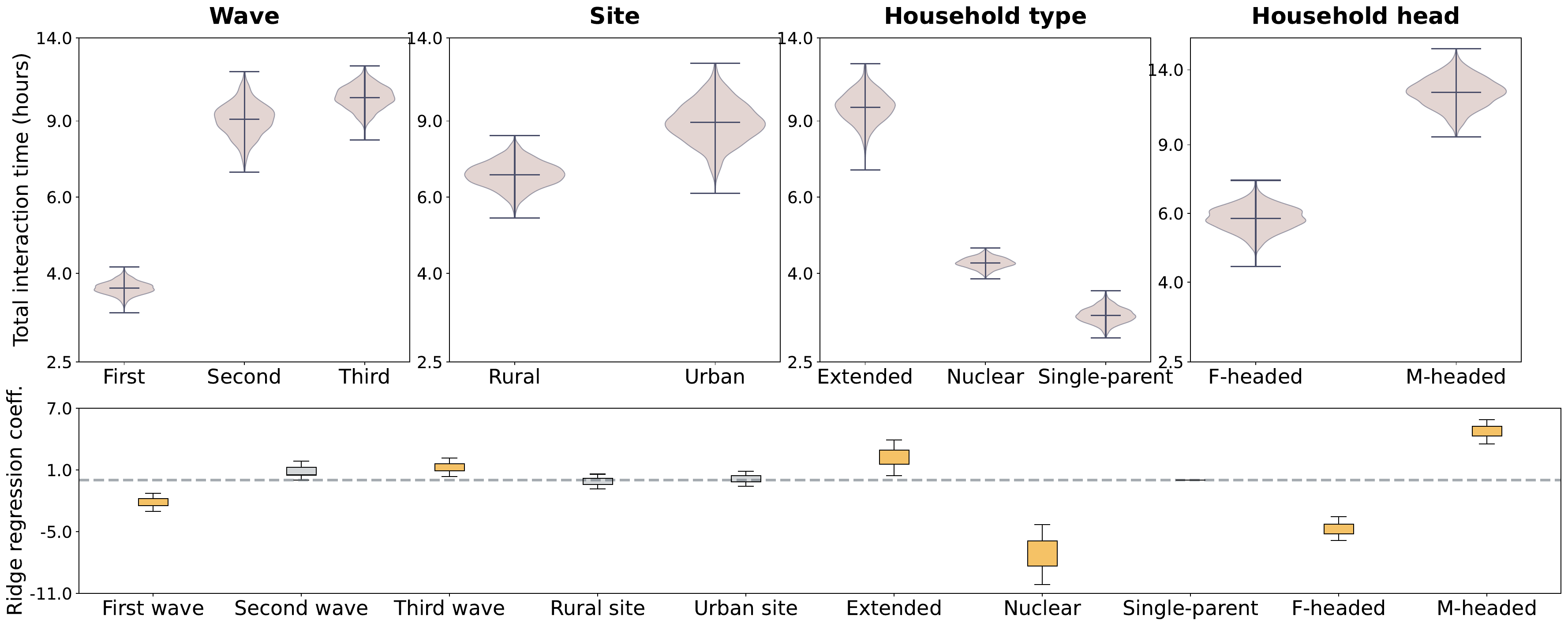}
	\caption{\textbf{Household interaction time per individual.} \textit{Top row: daily household interaction times normalized by the population, disaggregated by wave, site, household type, and gender of the household head.} The $y$-axis indicates the sum of all entries on the upper diagonal of the normalized household contact matrices shown in Figure~\ref{fig:HCM}. The distributions are obtained from a bootstrap sample, as described in Section \ref{sec:boot}. \emph{Bottom row: ridge regression.} Box plot of the ridge regression coefficients obtained from several bootstrap randomizations. The whiskers show the $5-95$ confidence interval. The face color is gray if the confidence interval crosses zero, and it is orange otherwise.}
	\label{fig:tot_time}
\end{figure}
\subsection{Comparison of the interaction rates}

We compare the interaction rates per individual for different household attributes. We consider the interaction rates per household, by age and gender, and with children. The analysis is performed on a bootstrapped sample of the empirical contact matrices, which is detailed in Section~\ref{sec:boot}. Size effects are quantified in terms of the Cohen's $d$ score, and larger scores imply higher separations between the distributions. To handle the dependencies between the household attributes (see Figure~\ref{tab:hhdistr}), we run a ridge regression to predict the interaction rates from the household attributes. This analysis allows us to identify statistically significant effects in predicting the interaction rates. The details of the ridge regression implementation are provided in Section~\ref{sec:ridge}. The statistical analysis is performed in Python, using the \texttt{numpy}, \texttt{pandas}, and \texttt{scipy} packages.

\subsubsection{Household interaction rate}
\label{sec:tot_time}

Figure~\ref{fig:tot_time} compares the distributions of the total interaction rate per household, defined as the sum of the upper-triangular elements of the matrices shown in Figure~\ref{fig:HCM}.

\medskip

The first wave (summer) recorded a much shorter average interaction time -- $3.7$ hours per individual on average -- with respect to the second (autumn) and the third (winter) -- $9.1$ and $10.2$ hours per individual on average, respectively. The size effect is much larger between the first and second waves ($d = 13$) than it is between the second and third ($d = 1.4$).
Higher average contact rates ($d = 3.1$) are observed in the urban setting -- $8.9$ hours per individual on average -- compared to the rural setting -- $6.8$ hours per individual on average.
Extended households have the highest average contact rates -- $9.8$ hours per individual on average --, followed by nuclear households -- $4.2$ hours per individual on average -- (size effect between extended and nuclear households $d = 14.4.$), and single-parent households -- $3.2$ hours per individual on average -- (size effect between nuclear and single-parent $d = 8.0$). 
Longer average contact rates ($d = 9.3$) are measured in male-headed households -- $12.3$ hours per individual  on average -- with respect to female-headed households -- $5.8$ hours per individual on average.

\medskip

The bottom frame in Figure~\ref{fig:tot_time} shows the results of the ridge regression analysis, which accounts for the dependencies between variables. This analysis confirms the statistical significance of the trends described above, except for the urban/rural classification, which is not significant. The observed difference in the second frame of the first row in Figure~\ref{fig:tot_time} can be explained by the high correlation between household headship (which has a significant impact) and the site location.

\begin{figure}[!t]
	\centering
	\includegraphics[width=\columnwidth]{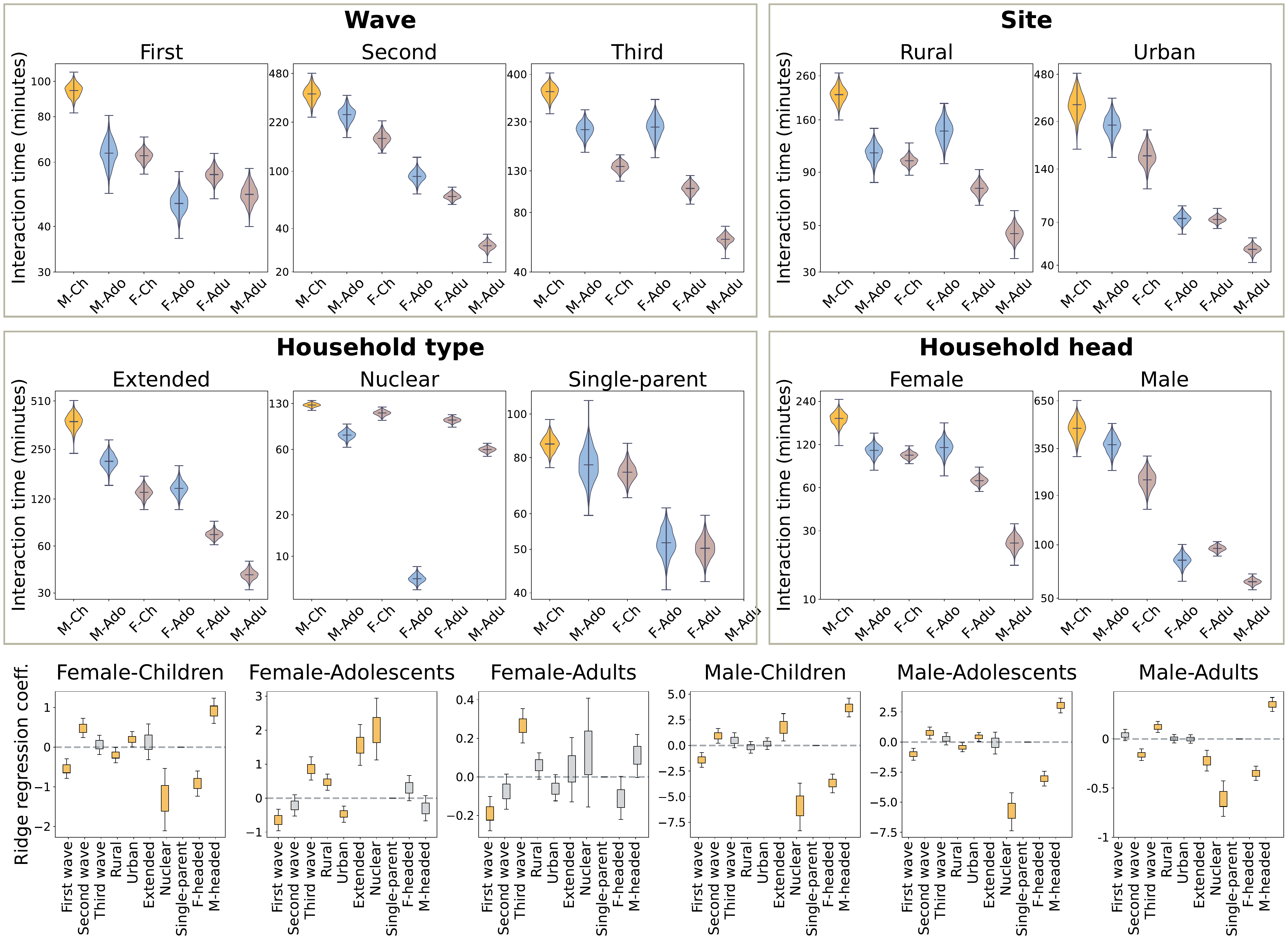}
	\caption{\textbf{Interaction rates disaggregated by age and gender.} \textit{Top row: daily interaction times by age and gender, normalized by the population and disaggregated by wave, site, household type, and gender of the household head.} The $y$-axis indicates the sum of the normalized household contact matrices rows. The distributions are obtained from a bootstrap sample, as described in Section~\ref{sec:boot}, and aggregated according to the measurement wave (first frame), the measurement site (second frame), the household type (third frame), and the gender of the household head (fourth frame). We highlight in orange the interactions involving male children \emph{M-Ch} and in light blue those involving adolescents \emph{M/F-Adu}. \emph{Bottom row: ridge regression.} Box plot of the ridge regression coefficients obtained from several bootstrap randomizations. The whiskers show the $5-95$ confidence interval. The face color is gray if the confidence interval crosses zero, and it is orange otherwise.}
	\label{fig:age_gender}
\end{figure}
\subsubsection{Interaction rates by age and gender}
\label{sec:age_gender}

We repeat the analysis of Section~\ref{sec:tot_time}, disaggregating the interaction times by age and gender. Figure~\ref{fig:age_gender} shows that, in all settings, male children experience the highest interaction rates. The smallest differences are recorded with male adolescents in the urban site ($d = 1.9$), in single-parent households ($d = 1.7$), and in male-headed households ($d = 1.9$). The involvement of female adolescents varies substantially in different contexts. It increases from summer to winter ($d = 7.6$ between first and second wave; $d = 6.9$ between second and third waves); it is larger in the rural context ($d = 5.5$), in extended ($d	 = 8.7$ with respect to single-parent households), and in female-headed households ($d = 3.4$). Adults, and in particular male adults, experience the lowest interaction rates in all contexts.

The ridge regression shows that men of all ages interact more often in male-headed households. Moreover, in extended households, female adolescents and male children experience higher interaction rates, while male adults have fewer contacts. Only female adolescents have more interactions in nuclear households.
\begin{figure}[!t]
	\centering
	\includegraphics[width=\columnwidth]{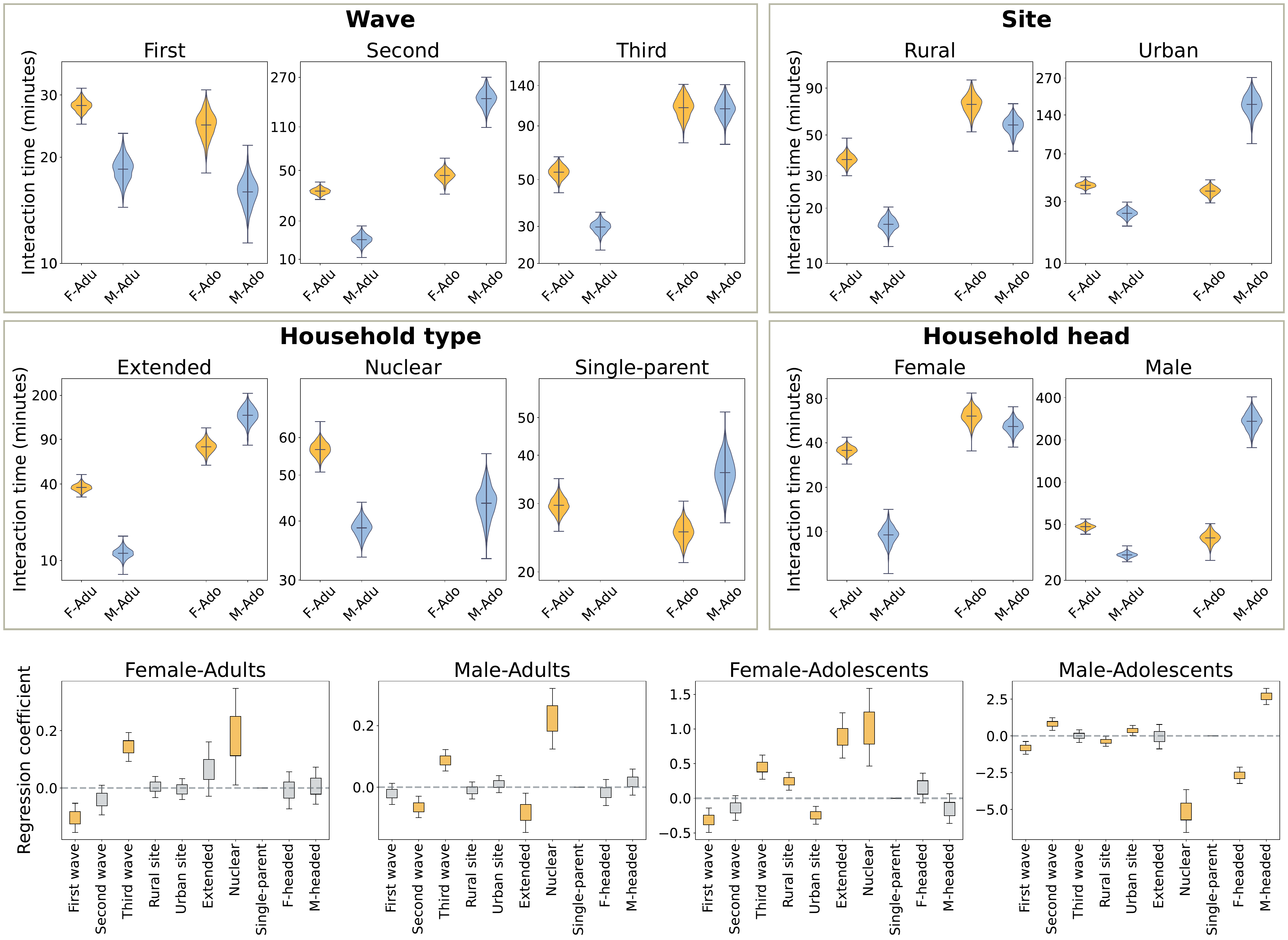}
	\caption{\textbf{Interactions with children.} \textit{Top row: daily interaction rates by age and gender normalized by the population, disaggregated by wave, site, household type, and gender of the household head.} The plot shows the histograms of the interaction rates with children (male and female) per individual, for adults and adolescents of both genders. The histograms are obtained from the bootstrap sample described in Section ~\ref{sec:boot} section and aggregated according to the measurement wave (top left frame), the measurement site (top right frame), the household type (bottom left frame), and the gender of the household head (bottom right frame). \emph{Bottom row: ridge regression.} Box plot of the ridge regression coefficients obtained from several bootstrap randomizations. The whiskers show the $5-95$ confidence interval. The face color is gray if the confidence interval crosses zero, and it is orange otherwise.}
	\label{fig:w_children}
\end{figure}

\subsubsection{Interactions with children}
\label{sec:w_children}

Figure~\ref{fig:w_children} shows that, in all contexts, female adults have higher interaction with children compared to male adults. This observation provides evidence that the burden of child care is mostly carried by women. The separation is smaller ($d = 10.2$) in nuclear households compared to extended households ($d = 14.6$), suggesting a more equitable share of child care between mothers and fathers in nuclear households. This is also confirmed by the ridge regression that identifies the ``Nuclear'' attribute as a positive predictor of the interaction rates between male adults and children. In our dataset, all single-parent households are female-headed and, therefore, cannot be compared.
Stark differences are observed in the interaction patterns of adolescents and children. As shown in Figure~\ref{fig:age_gender}, the involvement of adolescents drastically increases from winter to summer, and so does their interaction rates with children. Male adolescents generally have higher interaction rates with children, with few exceptions in which female adolescents record higher interaction rates: the rural households ($d = 2.4$), and female-headed households ($d = 1.5$). The regression shows that the ``Nuclear'' and the ``Extended'' household types positively predict interaction rates between female adolescents and children, while the ``Nuclear'' type negatively predicts interactions with male adolescents. Moreover, the gender of the household head is a statistically significant predictor of the interactions between male adolescents and children.

\subsection{Epidemiological impact of age and gender stratification according to the household classification}

Multi-class compartmental models, such as the SIR (Susceptible-Infected-Recovered) model, are widely used in epidemiology \cite{keeling2008modeling}. In this model, susceptible individuals ($S$) become infected ($I$) when entering into contact with already infected individuals. After some time, infected individuals recover ($R$). By including contact matrices in this model, one can account for across-group heterogeneities in the social mixing behavior. In this model, two parameters  -- $\beta$ and $\mu$ -- quantify the probability per unit time of infection and recovery, respectively. These parameters are disease-dependent and determine the basic reproductive number $R_0$, which is the average number of secondary infections caused by a single infected person in a fully susceptible population. Its value crucially determines the epidemic outcome since if $R_0 < 1$, the epidemic likely dies quickly, while if $R_0 > 1$, it is expected to spread rapidly through the population. For the SIR model, the $R_0$ has a known expression depending on the disease parameters $\beta, \mu$ and on the contact matrix $R$ \cite{diekmann2010construction}:
\begin{align}
	\label{eq:R0}
	R_0 = \frac{\beta\rho(RN^{-1})}{\mu}\,\,,
\end{align}
where $\rho$ denotes the largest eigenvalue of a matrix, and $N$ is the diagonal matrix containing the number of people in each group, specifically $N_{aa} = n_a$. For a given infectious disease, the parameters $\beta, \mu$ are fixed, and variations to $\rho(RN^{-1})$ across household groups imply differences in the epidemic outcome. As recently shown in \cite{manna2024generalized}, when combining multiple levels of stratification to define groups (for example, age-gender instead of only age), Equation~\eqref{eq:R0} still holds, but the spectral radius $\rho$ -- \emph{i.e.} the largest eigenvalue -- of the normalized contact matrix $RN^{-1}$ (and, consequently, $R_0$) cannot decrease. Therefore, single-attribute stratifications may be insufficient for capturing relevant heterogeneity patterns. 

\medskip

\begin{figure}[!t]
	\centering
	\includegraphics[width=\columnwidth]{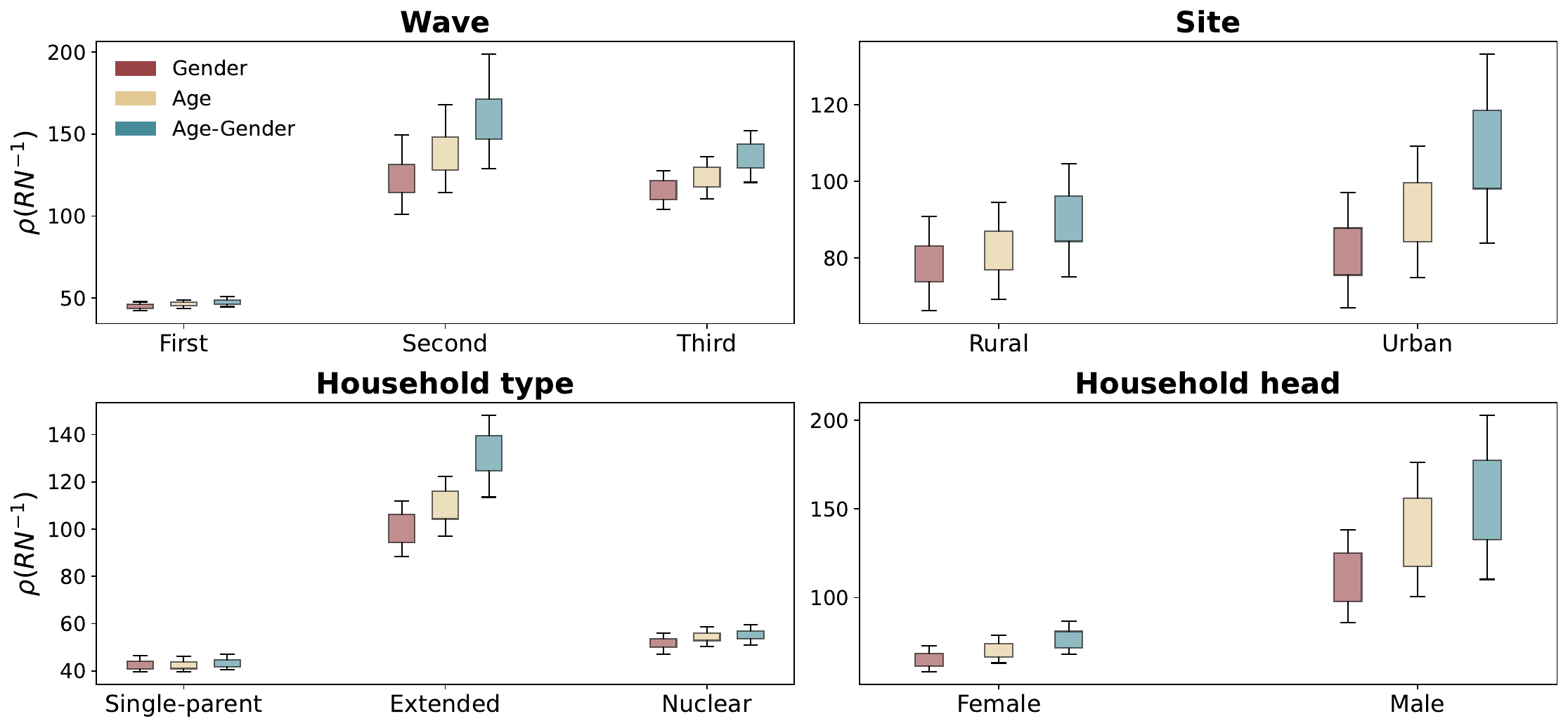}
	\caption{\textbf{Largest eigenvalue of the contact matrix for different aggregation and indexing strategies.} Each panel shows an aggregation based on a specific attribute indicated by the title. We consider three indexing strategies (shown on the x-axis) for each matrix: age, gender, and age-gender. The $y$-axis shows $\rho(RN^{-1})$, the largest eigenvalue of contact matrices the normalized contact matrix, which is proportional to $R_0$. The box plots are obtained for several bootstrap randomizations, as detailed in Section~\ref{sec:boot}. The whiskers indicate  the $5-95$ confidence interval.}
	\label{fig:R0}
\end{figure}
\begin{table}[!b]
	\centering
	\footnotesize
	\setlength\doublerulesep{0.2cm} 
	\begin{tabular}{|l|l|l|l|l|}
		\hline
		Wave & First & Second & Third\\
		\hline
		Gender & $1.52$ & $2.05$ & $2.34$ \\
		Age & $0.69$ & $1.14$ & $1.38$ \\
		\hline\hline
		Site & Urban & Rural & \\
		\hline 
		Gender & $1.44$ & $2.13$ & \\
		Age & $0.97$ & $1.25$ & \\
		\hline\hline
		Household type & Single-parent & Extended & Nuclear\\
		\hline
		Gender & $0.37$ & $3.31$ & $1.32$ \\
		Age & $0.39$ & $2.16$ & $0.32$ \\
		\hline\hline
		Household head & Female & Male & \\
		\hline
		Gender & $1.91$ & $2.19$& \\
		Age & $1.0$ & $0.83$ &\\
		\hline
	\end{tabular}
	\caption{\textbf{Effect size comparison between contact matrix specifications on the basic reproductive number $R_0$.} We compare three contact matrix types: ``Gender''-only (red in Figure~\ref{fig:R0}) ``Age''-only (yellow), and ``Age-Gender'' (blue). For each measurement wave, site, household type, and gender of household head, we compute $R_0$ using Equation~\eqref{eq:R0} with each matrix type. The table reports the $d$ score between ``Age-Gender'' and the two simpler specifications (``Gender''-only vs. ``Age-Gender'', and ``Age''-only vs. ``Age-Gender''). Higher $d'$ values indicate larger differences in the $R_0$ distributions.}
	\label{tab:r0}
\end{table}
The matrices shown in Figure~\ref{fig:HCM} utilize a two-level stratification that combines age and gender. To evaluate the significance of this stratification, Figure~\ref{fig:R0} compares the spectral radius of three contact matrices for each of the household classifications introduced in Section~\ref{sec:res}: one using only gender for indexing groups, another using only age, and a third combining both age and gender. The results are obtained for several bootstrap randomizations, as detailed in Section~\ref{sec:boot}. As expected, age accounts for a higher level of heterogeneity than gender, but less than the combination of both attributes. Table~\ref{tab:r0} reports the size effects between the distributions of $R_0$, comparing ``Gender''-only and ``Age’’-only contact matrices with ``Age-Gender'' matrices, testifying the importance of including both variables to explain the contact patterns in extended households where the complex patterns of interactions described in the previous sections lead to the largest increase in $R_0$ when using both variables at once.

\section{Discussion}

We conducted an analysis of the relation between in-household contacts and household attributes.
Our findings show the frequency of in-household interactions increases during winter, in agreement with other studies in Western countries that observed people spending more time outside the household during summer \cite{leech2002s, hussein2012activity}. Children, and in particular male children, are responsible for the highest contact rates, supporting the known fact that children are the primary introducers of influenza-like illnesses in households \cite{endo2019fine, ozella2018close}. However, accounting for the household composition is necessary to interpret the observed contact patterns better. Extended households are generally larger than nuclear and single-parent households and have higher interaction rates on average. Most of the activity involving male children is recorded in these households,  possibly due to the presence of siblings and cousins of similar age. In these households, adult men interact less than women (including adolescents). This observation agrees with the fact that in many African cultures, including South Africa, women are identified as the primary caregivers for children, even when they are not the biological mothers \cite{hatch2018cares}. In extended households, women are often in charge of caregiving also for the elderly \cite{chant2003female}. From an epidemiological perspective, these observations might provide support to the known fact that large and multigenerational households experience higher risks of transmission of diseases such as influenza, RSV, and measles \cite{endo2019fine, mahikul2019modeling, hilton2019incorporating, chisholm2020model}. 
In nuclear households, adult males experience higher interaction rates with children than in extended households. Fathers and mothers seem to undertake similar roles in child-rearing in this type of household. 
All single-parent households in our dataset are female-headed. This agrees with a global picture with $84\%$ of single-parent households run by women \cite{united2019progress}. The most common reasons explaining the gender gap are widowhood, divorce, the presence of children outside marriage, and migrant workers \cite{desai1998female}.

Female-headed households are more prevalent in rural settings and display a generally higher involvement of women. Nonetheless, we observed that male-headed have higher activity rates than female-headed households. This is a non-obvious result because female-headed households tend to be larger than male-headed households as they mostly fall under the extended category.

The differences between the rural and urban settings are mainly ascribed to a greater diffusion in rural settings of extended and female-headed households. The higher contact rates we observed in rural areas are consistent with previous results in Zimbabwe \cite{melegaro2017social}, but in contrast to other studies in similar contexts from low-resource settings in the sub-Saharan region \cite{kiti2014quantifying, del2021individual, dodd2016age, kleynhans2021cross, kiti2025social}, suggesting that the rural-urban context might not be as influential as the household composition or gender head. The higher involvement of women in rural contexts agrees with a societal norm that heavily influences household roles. The relatively better balance between gender roles in the urban setting can be attributed to relaxed societal gender norms \cite{pozarny2016gender}. Evidence from Zambia \cite{evans2014women} also suggests that in urban areas, caregiving responsibilities are becoming more balanced as men and male adolescents are increasingly involved in care duties, explaining the high involvement of males in interactions in the urban area. We remark that differences in results between our studies and other published studies could be, in part, as a result of different data collection methods.

We evaluated the epidemiological implications of observed differences in contact patterns by computing a proxy of the basic reproductive number associated with contact matrices collected in various social contexts and different seasons. The $R_0$ is mainly driven by age-based heterogeneities, even though the full degree of heterogeneity becomes apparent when accounting for both age and gender. Notably, the most significant variations are observed with changes in seasonality and in extended households, where complex relations are observed between age and gender attributes.

\medskip

Let us now comment on the main limitations of our work.
Our analysis is based on a small number of households. In some cases, only a few households are considered when disaggregating the data by site, type, and headship. To cope with this model and attribute statistical significance to our results, we performed bootstrap sampling. Given the small size of our dataset, however, our results are not representative of the population, and some household types (for instance, extended households) are over-represented in our dataset compared to the South-African population. Given the biases present in our dataset, our analysis cannot be interpreted as hypothesis-generating and conclusive. Moreover, the dataset used for this study contains only in-household interactions, and our conclusions do not apply to the across-household interactions that are fundamental in epidemic modeling. Despite these limitations, our results show the important role played by the household structure and composition in shaping face-to-face interactions. These results contribute to addressing the limited data gap on face-to-face interactions in sub-Saharan Africa, where the prevalence and re-emergence of infectious diseases pose a significant public health challenge. Our results encourage future data collection on face-to-face interactions to include rich metadata information related to the social role of individuals in the household.

\section{Methods}
\label{sec:methods}

\subsection{Ethics}
\label{sec:ethics}

Ethics permission to conduct the experiment was received from the Wits Human Research Ethics Committee (Medical) (ethics reference no. 150808). The project and its implementation were also discussed with community advisory boards at both study sites. The study was designed and implemented by South Africans. In each study site, the collaborators implementing the study had a long-standing relationship with the community. Informed consent documents were translated into the local languages and discussed with community advisory boards. Written informed consent/assent was obtained from all study participants.

\subsection{Data description}
\label{sec:data}

The PHIRST study \cite{cohen2021cohort, kleynhans2021cross, dall2022estimating} was a perspective household cohort study, conducted in South Africa during $2016-2018$ to investigate the burden and transmission dynamics of influenza and respiratory syncytial virus (RSV). Randomly selected households were approached for enrolment, and households with  more than $2$ members and where at least $80\%$ of household members consented, were included in the study. A new cohort was enrolled each year, with enrolment occurring in the November through to December the prior year, and intensive follow up occurring from February - October with twice weekly home visits and nasopharyngeal swab collection for testing of influenza and (RSV), irrespective of symptoms. Recruitment occurred at the end of the preceding year to ensure all households were enrolled before the start of the RSV season which starts in February/March in South Africa.  A household contact survey was nested in PHIRST in 2018, in which all households included in the 2018 cohort were approached to partake. During this data collections, contact patterns were measured with the proximity sensors developed by the \texttt{SocioPatterns} collaboration (\href{http://www.sociopatterns.org/}{sociopatterns.org}, \cite{cattuto2010dynamics}). These sensors are non-obtrusive devices worn on the chest of the study participants during the data collection. They emit low-power radio signals, exchanging approximately $40$ information packets per second when facing one another. They record a proximity interaction if two sensors exchange at least one packet in a $20$-second period. In this case, each sensor records the unique identifier of another interacting sensor, the interaction time, and the power attenuation from the sender to the receiver. Proximity interactions within approximately $1.5$ meters are identified by filtering the measurements based on their signal attenuation. These sensors provide high-resolution measurements of human interactions and have been deployed in several contexts, including schools \cite{stehle2011high, fournet2014contact}, hospitals \cite{vanhems2013estimating}, and conferences \cite{genois2018can}, among others. The proximity sensors were packaged in PVC pouches and worn as a badge pinned to either the inside or outside of the clothing with a safety pin, or on a lanyard around the neck, depending on the preference of the participant. Field workers visited the household at least two times during the deployment and reminded participants to wear the sensors. We followed the data cleaning procedure described in \cite{dall2022estimating}, which we here briefly summarize. 
	\begin{enumerate}
		\item Only interactions between members of the same households are kept, as they are the only ones needed to define household contact matrices.
		\item The field team performing the data collection reported the dates of beginning and end of each deployment for every household. All the interactions recorded beyond the official measurement window are dropped. The compliance with the experiment protocol was extremely variable in those days, making these measurements unreliable. 
		\item We removed the interactions occurring during the first and the last days of each deployment. During these days, we observed anomalous activity patterns due to the sensors' dispatching.
		\item Some households were excluded because of the impossibility of correctly mapping the sensor data with some (or all) of the household members.
		\item All households with fewer than days of valid measurements per wave were excluded. We recall that in each measurement wave we collected data for at least $10$ days.
		\item Households with non-circadian activity patterns (amenable to a misuse of the sensors) were excluded.
\end{enumerate}
Our cleaned dataset is composed of $307$ individuals and $60$ households.

\subsection{Household classification}
\label{sec:taxa}

Our results are based on a household classification made according to the site in which household are located, its headship and its type. Below we provide additional details describing and motivating this classification.
\begin{itemize}
	\item \textbf{Site (rural or urban).} Socio-demographic characteristics of individuals and households differ across rural and urban settings. Hence, it is important to capture differences and similarities in contact patterns driven by these attributes. The rural site is composed of $28$ households, and the urban by $32$. 
	\item \textbf{Gender of the household head.} The household head does not have a unique definition, but it generally refers to the person who holds the economic responsibility (for example being the primary breadwinner) or who has the most decision-making power in the household \cite{chant2011household}. In our study, the household head is the person identified by the household members as the primary decision-maker. Especially in South Africa, this role is commonly held by the oldest individuals who typically also have the highest income \cite{posel2001heads}. Regardless of the actual dynamics within the household, men are commonly more likely to be designated as household heads \cite{chant2011household}, while women-headed households are more susceptible to economic hardships \cite{posel2001heads}. Our dataset comprises $43$ female-headed households ($24$ in Agincourt, $19$ in Klerksdorp) and $17$ male-headed households ($4$ in Agincourt and $13$ in Klerksdorp). This agrees with the general trend in South Africa that sees female-headed households more prevalent in rural areas \cite{dungumaro2008gender}.
	\item \textbf{Household type.} For this classification, we follow \cite{ibisomi201410}, focusing only on the household types observed in our cohort. These include: (i) \emph{nuclear households} ($12$ households), formed by a couple (married or cohabiting) and their biological or adopted children; (ii) \emph{single-parent} ($13$ households) in which the household head lives with his/her children; (iii) \emph{extended households} ($33$ households), composed by relatives beyond nuclear family such as cousins, aunts, uncles, and grandparents. Our dataset further comprises two households classified as \emph{complex} as they also contain non-relative members. Given that only two households belong to this class, we chose to absorb them into the extended and nuclear households, according to their composition. In our dataset, extended households are larger on average with $5.7$ individuals against $3.9$ and $4.2$ observed in nuclear and single-parent households, respectively.	According to the General Household Survey \cite{StatsSA2022}, nuclear households are the most common in South Africa, reaching $40.1\%$ of the total. In our dataset, extended households are over-represented in both contexts: $57\%$ in Agincourt versus $45\%$ observed in rural areas; $53\%$ in Klerksdorp versus $30\%$ observed in urban areas. On the opposite, nuclear households are less represented than expected but, in agreement with the national trend, they are more frequent in the urban area ($28\%$ in Klerksdorp) than in the rural one ($11\%$ in Agincourt).
\end{itemize}

\subsection{Contact matrices definition}
\label{sec:boot}

For each household $h$ in the dataset, and a pair of age-gender indices $a, b$ (for instance $a = $ male-children, $b = $ female adults), we denote with $R_{ab}^{(h)}$ the daily average interaction duration aggregated between the groups $a$ and $b$. We further let $n_a^{(h)}$ be the number of individuals in age-gender group $a$ in household $h$.
By letting $\alpha$ be one of the household attributes described in Section~\ref{sec:data} (\emph{e.g.} ``male-headed'', or ``rural''), we denote with $\chi_{\alpha}$ a random sample with repetition of $1000$ households having the attribute $\alpha$. The corresponding household contact matrix (HCM) is defined by the following equation
\begin{align}
	\label{eq:CM}
	\tilde{R}_{a,b}(\chi_{\alpha}) = \frac{\langle R_{ab}\rangle_{\chi_{\alpha}}}{\langle n_a\rangle_{\chi_{\alpha}}\langle n_b\rangle_{\chi_{\alpha}}}\,\,,
\end{align}
where $\langle M \rangle_{\chi_{\alpha}} = \frac{1}{|\mathcal{\chi}_{\alpha}|}\sum_{h \in \mathcal{\chi_{\alpha}}} M^{(h)}$ is the empirical average of the attribute $M$ over the set $\chi_{\alpha}$. For instance, let $\alpha$ denote male-headed households, then the matrix $\tilde{R}(\chi_\alpha)$ is the empirical average of all household contact matrices collected in male-headed households, and normalized by the average population sizes in these households.
Note that each household can appear more than once in the set $\chi_{\alpha}$. The bootstrap distribution is obtained over $1000$ random samples of the set $\chi_{\alpha}$.

\subsection{Ridge regression}
\label{sec:ridge}

To disentangle the dependencies of the household attributes in shaping the interaction patterns, we run a ridge regression in which we predict the contact rates from a list of household attributes. First, we define a sample $\chi_i$ of $100$ (households-deployment) pairs drawn with repetitions from the whole dataset, composed of $180$ instances. Using Equation~\eqref{eq:CM}, we obtain the corresponding matrix $\tilde{R}_{\chi_i}$. We then extract a scalar quantity $y_i$ from $R_{\chi_i}$ (\emph{e.g.}, the total interaction time), that needs to be predicted from the regression model. Finally, we define a feature vector $\bm{x}_i$ whose entries indicate the number of sampled instances per attribute. The attributes are first, second, and third wave, nuclear, single-parent, and extended households, male and female-headed households. By repeating the process for different randomizations $i = 1,\dots, 1000$, we obtain a target vector $\bm{y}$ and a feature matrix $X$ that we use as inputs for the ridge regression. We then repeat the process $1000$ times and obtain a distribution of the fitted parameters, which we consider to be significant only if the $5-95$ confidence interval does not cross zero.

\section*{Funding}

This work was supported by the National Institute for Communicable Diseases of the National Health Laboratory Service and the US Centers for Disease Control and Prevention (cooperative agreement number 5U51IP000155). KT acknowledges funding received from the European Union’s Horizon 2020 Research and Innovation Programme under the Marie Skłodowska-Curie Evogames project grant agreement number 955708. LD, CCa, and DP acknowledge the support from the Lagrange project of Fondazione CRT. The funders had no role in study design, data collection and analysis, decision to publish, or preparation of the manuscript.

\section*{Competing interests}

CCo has received grant support from Sanofi Pasteur, US CDC, Welcome Trust, Taskforce for Global Health and Bill \& Melinda Gates Foundation.

\section*{Data availability}

Data and codes to reproduce the results discussed in this paper are available at the following public repository: \href{https://github.com/lorenzodallamico/PHIRST_SocioPatterns}{github.com/lorenzodallamico/PHIRST\_SocioPatterns}.


%

\end{document}